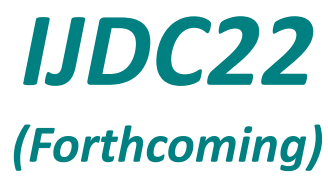

# Selecting efficient and reliable preservation strategies: modeling long-term information integrity using large-scale hierarchical discrete event simulation

*(Research Paper)*

Micah Altman
Massachusetts Institute of Technology

Richard Landau
Massachusetts Institute of Technology

**Abstract**

This article addresses the problem of formulating efficient and reliable operational preservation policies that ensure bit-level information integrity over long periods, and in the presence of a diverse range of real-world technical, legal, organizational, and economic threats. We develop a systematic, quantitative prediction framework that combines formal modeling, discrete-event-based simulation, hierarchical modeling, and then use empirically calibrated sensitivity analysis to identify effective strategies.

Specifically, the framework formally defines an objective function for preservation that maps a set of preservation policies and a risk profile to a set of preservation costs, and an expected collection loss distribution. In this framework, a curator's objective is to select optimal policies that minimize expected loss subject to budget constraints. To estimate preservation loss under different policy conditions optimal policies, we develop a statistical hierarchical risk model that includes four sources of risk: the storage hardware; the physical environment; the curating institution; and the global environment. We then employ a general discrete event-based simulation framework to evaluate the expected loss and the cost of employing varying preservation strategies under specific parameterization of risks.

The framework offers flexibility for the modeling of a wide range of preservation policies and threats. Since this framework is open source and easily deployed in a cloud computing environment, it can be

Correspondence should be addressed to Micah Altman, 6104 Building NE36, MIT, Cambridge, MA, 02139
Email: <micah.altman@gmal.com>

Repository of source code available at https://github.com/MIT-Informatics/PreservationSimulation

used to produce analysis based on independent estimates of scenario-specific costs, reliability, and risks.

We present results summarizing hundreds of thousands of simulations using this framework. This exploratory analysis points to a number of robust and broadly applicable preservation strategies, provides novel insights into specific preservation tactics, and provides evidence that challenges received wisdom.

# Significance

Deploying cost-effective, reliable bit-level long-term preservation at scale remains an unsolved problem. [Rosenthal 2010; Altman *et* al. 2015, 2020] Memory organizations have identified a number of high-level 'best practices', such as fixity checking and geographically distributed replication, but there is little specific guidance or empirically-based information on selecting specific preservation strategies that fit a curating institution's risk-tolerance, threat profile, and budget. Thus, while cloud storage vendors such as Amazon tout 99.999999999% durability; these claims typically lack substantial explanation or even clear definitions [see e.g. Mellor 2018] Further, professional memory organizations vary significantly in the practices they use, and how they use them -- even in the number of copies held. [Gallinger *et* al. 2017]

Strategies for preserving digital information are generally based on the observation that neither digital media, nor formats, nor institutions are reliably durable. While a number of 'good practices' are recognized for digital preservation, many of these practices are heuristic, and most are based on experience with particular technologies and threats. Stewards of digital information are faced with a large set of choices in developing a preservation strategy. These choices include document size and data format; file encryption and compression; storage media durability and reliability; collection replication, distribution, verification, and repair. [see e.g. Gallinger *et* al. 2017] These choices have the potential to change dramatically the cost of a preservation strategy, and how (and where) that strategy is vulnerable to a wide range of threats. Moreover, changes in these factors interact in complex ways, making it difficult to discover optimal/efficient strategies.

This article addresses the problem of formulating simple, efficient, and reliable operational preservation policies that ensure bit-level information integrity over long periods, and in the presence of a diverse range of real-world technical, legal, organizational, and economic threats. We develop a systematic, quantitative prediction framework that combines formal modeling, discrete-event-based simulation, hierarchical modeling, And then use empirically calibrated sensitivity analysis to identify effective strategies.

# Methodology

The ultimate goal of information preservation is to communicate across time. Our concrete objective, broadly speaking, is to maintain a collection of documents, so that its contents can be read at a designated future time. Communication will be deemed a success if at some designated future time the integrity of the documents has been maintained. (In the full paper we extend this to the case where additional context about formats, encryption, etc. must be preserved so that the document can be meaningfully rendered.)

More formally, we model the curator's task as the selection of preservation practices and parameters (e.g., auditing frequency) based on feasible practices and available systems, such that preservation costs are minimized, such that the expected loss of content does not exceed the curator's target given a specified risk profile. (In the full paper, we also model a dual problem of minimizing loss given a fixed budget.)

Curators might wish to be aware of the technologies and concerns in the areas we cannot control. In defining the curatorial strategy, we focus on those elements that curators are likely to readily control [Gallinger, *et* al. 2017] : the number and distribution of copies, how to audit and repair these, and whether to apply file transformations such as compression or encryption.

Prior work by [Baker *et* al. 2006], which we extend, uses single-level simulation to examine a core tradeoff between replication and auditing. Others have used related approaches -- such as Markov Chain Simulation [Lebrecht *et* al. 2011; Li *et* al. 2012] to estimate data failure at the hardware storage layer under simplified threat models in relation to the choice of storage approaches (e.g. RAID configuration). Work such as [Pinheiro et al. 2007] summarizes empirical rates of storage failures -- we use this and subsequent work to calibrate the model we present.

Where both cost and loss functions are simple and behaved, it may be possible to find the optimal solution through closed-form mathematical analysis or simple Monte-Carlo simulation. However, in more realistic conditions, risk of loss is a complex function of multiple threats at multiple levels [see, e.g. Rosenthal, et al. 2005] -- including low-level media failures, mid-level events such as manufacturing defects that affect clusters of media, and high-level events such as government action that can simultaneously affect multiple replicas of entire collections. Thus we use computationally-intensive discrete event simulation to estimate the losses under different proposed strategies.

Our underlying storage model is also hierarchical. A client (library) has a collection of documents in digital form. These documents are recorded on sectors of a storage medium; errors in the sectors within it cause a document to be lost -- as illustrated in **Figure 1**. As also illustrated in Figure 1, large documents are larger "targets" than small documents and are therefore more likely to be hit by random errors.

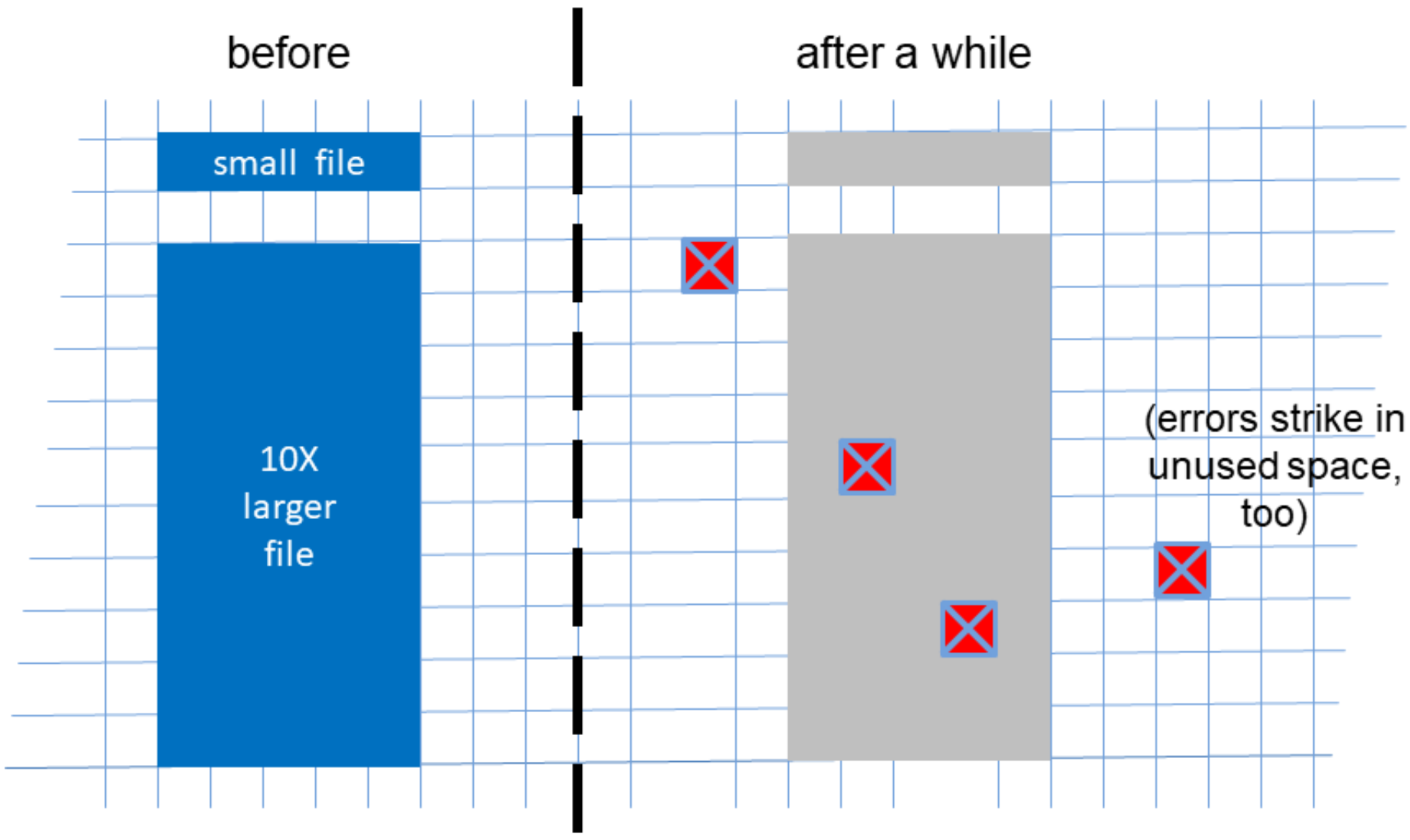

**Figure 1: Documents are stored on sectors of disk (or other) storage. Large files occupy more sectors than small files. When errors occur randomly in sectors, large files present a larger "target area" and are more likely to be hit by an error.**

In the core simulation model, we treat a document with any sector errors as corrupt, and use a document size that is constant number of sectors**.** Although these may seem like rigid assumptions, it turns out that the predictions of these model can be transformed through a simple linear formula to apply to larger and/or more robust documents. We discuss computing loss for different document sizes in Appendix A, and robust documents in the results section under *Modeling Robust Documents and Compression Risks.*

A copy of the collection of documents is stored on a server maintained by a separate institution. If the client maintains multiple copies of the collection of documents, then several copies are stored on separate, independent servers. Customers retrieve documents from the server(s) to read them. An error may occur that corrupts a portion of a document or makes that copy inaccessible. In this case, we consider the copy to be lost. Other copies may still persist. If all copies of a document are lost, then the document itself is permanently lost.

Specifically, we are concerned with four hierarchical levels of failures that affect document collections.

- **Disk Sector Failures.** Small scale errors in disk recording can result in partial or total failure of a document copy. For the purposes of this study, we have assumed that a single failure in the body of a document's data causes the document to be considered lost.
- **Environmental Glitches.** The rate at which errors arrive in disk sectors is not always constant. Transient environmental conditions, "glitches," such as failures of HVAC, electrical noise, and such can raise the rates of sector errors.
- **Server Failures.** Data servers are not immortal. Companies may go out of business, be bought by or merge with other companies, change their mission, and so forth. Our research covers a range of expected lifetimes of server companies.
- **Major Shocks.** Economic and regional environmental conditions can cause multiple servers to fail in short periods of time. Economic downturns can stress many companies in an industry. Regional environmental disasters such as earthquakes, floods, wars, and such can result in failure, or inaccessibility, of multiple servers. And government censorship can make servers or groups of servers inaccessible and lost to a replication set of servers. Such correlated failures are a serious threat to a collection and require more intense monitoring by the client library to ensure the health of the collection.

Note that these distributions imply some assumptions about operations:

- The model assumes that some form of local error-correcting storage, such as RAID is used, and that the benefits of server-local error correction, such as the use of RAID storage be incorporated in the logical sector error rate. Thus the sector error rate should reflect the ex-post rate of sector failures, after any RAID repairs. This rate effect of different RAID configurations, pattern of single disk failure and subsequent repair, etc. are not directly modeled.
- The model assumes that a storage unit is replaced, and content automatically migrated, after its normal service life. If not, sector failure risk would typically accelerate after the 3-5 year service lifetime of the initial hardware. We can still model failure to migrate media modeled through introduction of environmental glitches that increase sector error failure.
- The model assumes that the failure of institutions is conditionally independent, absent shocks. In other words, when two servers share an internal dependency -- such as reliance on the same third-party storage layer, a shock should also be added to the model to reflect that dependency.

In Table 1 we summarize these failure types, their distribution, and their visibility. We also provide examples of real world failures that one can represent using these event types.

| | Characteristics | | | Exemplar Threats | |
|---|---|---|---|---|---|
| Layer | Role | Visibility | Distribution | Lower Frequency | Higher Frequency (lower severity) |
| Hardware (Sector) | Causes sector error / single document loss | Silent. [1] | Poisson event | Controller failure | Media corruption. |
| Local environment (Glitch) | Increases rate of storage error | Invisible[2]. | Poisson event; triggers environmental change of some duration | HVAC failure | Power spikes |
| Institution (Server Failure) | Causes loss of a single copy of a collection | Silent. | Exponential Lifetime | Ransomware Business failure | Curator error. Billing error |
| Macro Environment (Major Shock) | Increases rate of server failure | Invisible. | Poisson duration | Corporate Mergers | Recession |
| | Immediate loss of multiple servers | Silent or visible | Poisson event | Government Suppression | Regional war |

**Table 1: A Hierarchical Typology of Preservation Threats. Small errors corrupt storage sectors of individual documents. Greater threats cause entire server(s) to fail, losing all copies of documents on the server(s).**

Note that these distributions imply some assumptions about operations:

- The model assumes that some form of local error-correcting storage, such as RAID is used, and that the benefits of server-local error correction, such as the use of RAID storage be incorporated in the logical sector error rate. Thus the sector error rate should reflect the ex-post rate of sector failures, after any RAID repairs. This rate effect of different RAID configurations, pattern of single disk failure and subsequent repair, etc. are not directly modeled.
- The model assumes that a storage unit is replaced, and content automatically migrated, after its normal service life. If not, sector failure risk would typically accelerate after the 3-5 year service lifetime of the initial hardware. We can still model failure to migrate media modeled through introduction of environmental glitches that increase sector error failure.
- The model assumes that the failure of institutions is conditionally independent, absent shocks. In other words, when two servers share an internal dependency -- such as reliance on the same third-party storage layer, a shock should also be added to the model to reflect that dependency.

[1] Detected on audit

[2] Inferred through indirect effects on other error

**A Simple Cost Model**

Many storage vendors may be available to a client, each with charge schedules. For the most part, vendors will charge for storage and bandwidth.

- A charge per month per byte stored (usually per gigabyte or petabyte). The cost of storage may vary by "quality" of storage, based on its typical error rate or perhaps on speed of retrieval access. Storage is charged per copy; multiple copies cost more.
- A charge per month per byte sent in or out ("ingress" and "egress" charges). Bytes sent do not distinguish between user access for normal retrieval and administrative access for auditing. The cost may vary by speed or reserved bandwidth (Mbps).
- Charge schedules for storage and bandwidth may include quantity discounts.

For the purposes of this study, a client will store a collection on a set of servers of the same "quality" level. Documents with differing quality requirements are considered separate collections and are stored and managed separately.

# Results

In this section we show how simulation of the multi-level failure model can be used to design a robust preservation strategy: First, we start by modeling sector-level errors that cover a very wide range of error rates, more than three orders of magnitude. We find that maintaining multiple copies of a document collection, along with a regimen of regular auditing, can preserve the collection over a range of error rates wider than is likely to be encountered with real commercial disk drives. Second, we introduce small and large glitches, which increase the base error rate from two to ten times. We find that such temporary excursions are indistinguishable from minor or even major differences in the base error rate, and that the prior strategy remains robust. Third, we introduce whole server failure. We find that, by incrementally increasing the number of copies and auditing frequency, a client library can protect the collection against individual server failures over a wide range of server lifetimes. Finally, we introduce major shocks that increase the rate of server failure and/or simultaneously eliminate up to three servers. We find that sufficiently increasing frequency of auditing and repair can protect a collection against even the shocks induced by major recessions and minor wars.

**Constructing a Preservation Strategy that is Robust to Base Storage Quality**

If a collection exists as only a single copy, then it is very likely that some of its documents will be lost within a decade, even if the storage medium is highly reliable. **Figure 2** shows the likelihood that documents will be lost, over a decade or longer, depending on the sector lifetime of the storage medium. If there is only one copy of a document, then any error causes permanent loss of that document. If there are multiple copies, then some undamaged copies may remain. But the number of (unaudited) copies required to prevent permanent losses is very large.

Further, even if the rate of error accumulation is much lower than illustrated above, many documents will be lost over longer periods of time: For example, over 50 years, about 20 parts per million (ppm) to 200ppm of the collection will be lost, depending on document sizes, even if media reliability were 10x the maximum shown. [3]

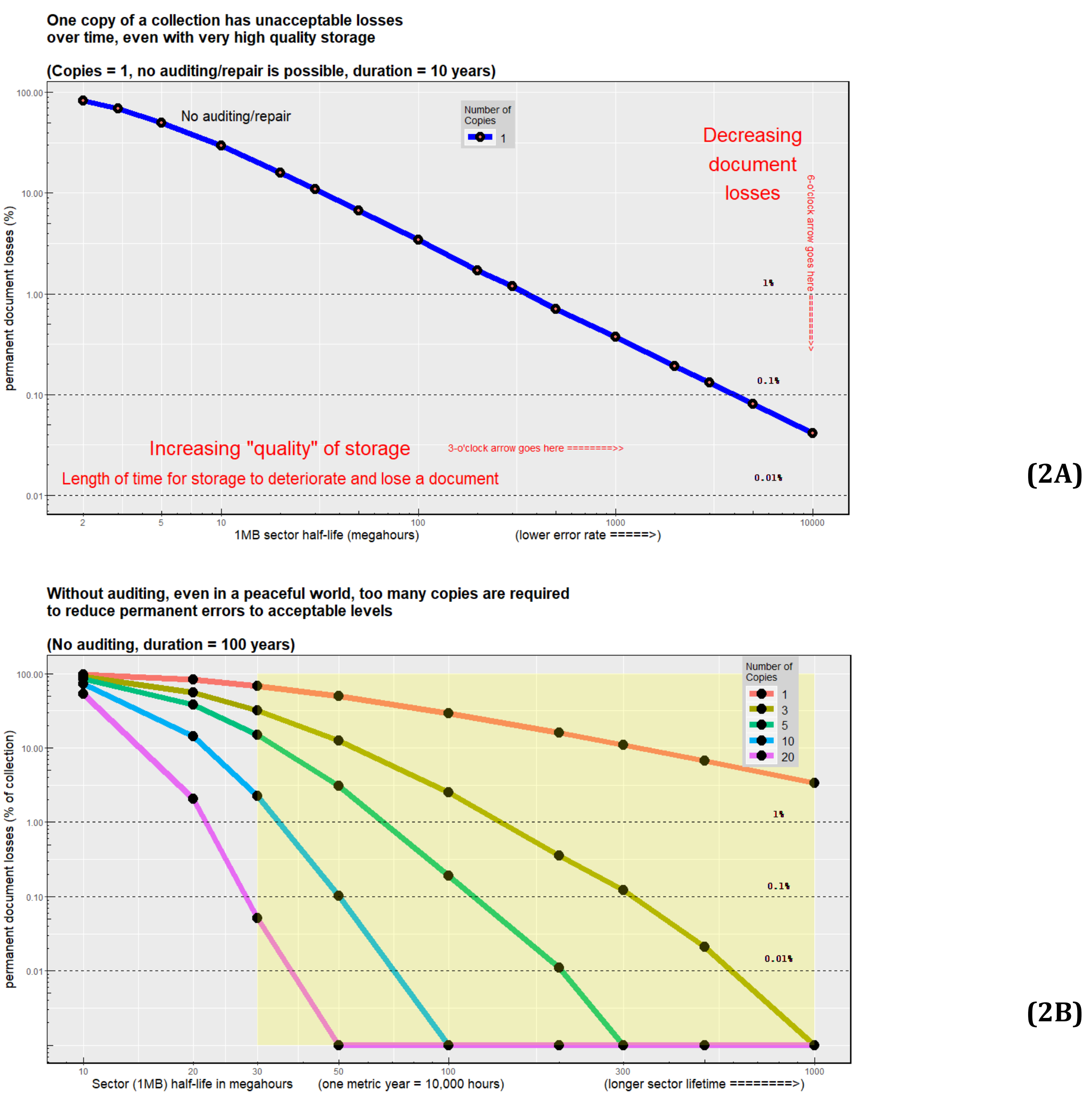

**Figure 2. (2A) A single copy of a document collection suffers considerable losses over a short period of time, even with a very high quality storage medium. See Appendix B for discussion of the calibration of this figure. (2B) Over long periods, even very large numbers of unaudited copies will suffer permanent losses. The shaded area covers what we consider to be the plausible region of disk quality.**

[3] In the graphs, very small numbers are grouped together. All values in that range are in fact zero. True zeroes are biased away from the origin by a few ppm to permit the use of logarithmic axes.

A single copy of a document is vulnerable, regardless of the quality of its storage medium. Multiple copies of collections are less vulnerable but still deteriorate over long periods, as sometimes random errors will coincide to cause a permanent loss where all copies of a document have been corrupted. Simulations demonstrate that, without regular auditing and repair, no *reasonable* number of copies will prevent significant document losses over a long period See **Figure 3**, which shows expected losses with various numbers of multiple unaudited copies.

In addition, we simulated environmental disturbances such as variations in temperature, humidity, dust, etc., due to HVAC or electrical problems that act to accelerate sector failure for a period of time. We find that the effect of such "glitches" has the same effect as directly increasing the base error rate. Moreover, this similarity implies that preservation strategies (such as those we recommend below) that are robust across a range of sector failure rates are also robust to moderate to severe glitches that increase the sector error rate by factors of three to ten times.

**Estimating a range of deterioration rates of sectors**

At what rate do disk sectors deteriorate and lose document information stored on them? Very little direct information is available from manufacturers or from large disk consumers such as cloud storage facilities regarding either theoretical or empirically measured error rates at the sector, file or collection levels. Most published claims regarding storage reliability are either so ill-defined as to be unmeasurable (see e.g., Mellor 2018 on AWS and Azure claims of "sixteen nines" of reliability), entirely theoretical (see. e.g. Rosenthal 2010, on the calculation of mean-time-to-data-loss by storage manufacturers), or measure the failure of entire drives during a service lifetime. Based on these latter, evidence-based estimates, hard-drives fail at the rate of roughly 1.25-4% annually (see Backblaze 2018) -- when deployed professionally.[4]

If no redundancy (e.g. RAID, erasure codes) is deployed at the filesystem level, the observed disk failure rate would be an upper-limit on sector life as well -- if the disk cannot be read, no sector within it could be read. Based on the lowest observed rate of 1.2% annually, and typical drive sizes of 1-2TB, the implied maximum sector half-life of a non-RAIDED system would be 250,000KH. However, this number is implausibly *low* in an environment in which some file-system redundancy is used -- at this level we would observe frequent losses of large files. For realism, we assume that RAID or other filesystem is deployed effectively to the extent that the system can be run in production without obvious failures.

Closed-form calculations based on the assumption of Poisson arrivals of sector errors can be used to approximate sector reliability based on experience with small storage systems.[5] (Cf Appendix B.) How high a sector failure rate would the average computer user tolerate on, say, a system disk? Under a reasonable set of assumptions about document size and sector size, we find that half-lives for megabyte-size disk regions less than about 100 megahours (100E06 hours) would result in a noticeable proportion of files being lost in the first few years of operation, for example, more than two percent of documents in just the first two or three years of disk use. Such losses would render an operating system disk unusable, and would certainly be too high for a data (document) archive. At the other end of the reliability spectrum, sector half-lives more than approximately 1,000 megahours would characterize disks that were nearly perfect and very long-lasting. In the absence of sound empirical data on this topic, we consider the range of sector half-life from 100 megahours to 1,000 megahours to be a plausible range for commercial disk drives that corresponds to real-life experience. However, due to the possibility of environmental glitches (failures of air conditioning, noisy electrical power, etc.) that increase error rates, we would extend the realistic range down to 20 or 30 megahours, as shown in the shaded region of Figure 2B. We have extended the range of values for simulations to include these less optimistic values.

[4] Electronics generally experience accelerated failure rates during their initial burn-in period, and after their service lifetime (see e.g. Xin, et al 2003). And most hard drives are designed for a service life-time of 3-5 years.

[5] To simplify calculations and to make our results more accessible, we have simplified some of the numbers and units in the simulations and graphs. Error rates for disk sectors and for servers are stated in half-lives in units of kilo-hours and mega-hours. Also, simulation event time periods are stated in "metric years" of 10,000 hours. We feel that this change in the length of a year makes the results slightly more conservative, since the simulation's year and ten-year periods are somewhat longer than calendar years. Also, to save compute cycles in simulations, the collection size is set to 10,000 documents; results are easily extrapolated to larger collections.

**The Need For Auditing**

We must adopt active auditing strategies to detect and correct errors in data in order to preserve the corpus over long periods. Such strategies must include three components: detecting errors by fixity information or other means; correcting errors, usually by replacing a damaged copy with a known good copy; and actively locating errors by patrolling through the data to examine all the documents. (Baker, et al 2006)

We need to be wary that failures -- of documents or servers -- are *silent* to the client. A client cannot afford to wait until a document is requested to discover that all copies have been lost. An auditing system must actively search through the data for latent errors in order to locate (and repair) them before these errors pile up and overwhelm the redundancy of storage.

An auditing cycle may be accomplished in a single pass through the documents or broken up into several “segments.” It is important to note that total auditing requires that all copies of a document be checked during each auditing cycle. A document may be assigned to any segment within a cycle, but it must be present in some segment of each cycle. The sampling of documents for each segment of the auditing can be systematic or random; but it is important that the auditing actually examine *all* documents. That is, auditing segments must sample documents *without replacement* each cycle. Sampling with replacement permits some documents to be missed in each cycle and reduces the effectiveness of auditing.

Random auditing is often expressed as, for instance, "audit ten percent of the documents every month." The difference between this random strategy and segmented auditing is that the random selection may be chosen *with replacement*. Thus it is likely that some documents will escape auditing during each cycle. This observation is analogous to the experiment of throwing a thousand balls randomly into a thousand urns. Since the balls arrive according to a Poisson distribution, some of the urns will receive no balls, some one, and some more than one, according to distribution. Documents that are audited zero times are not being audited effectively at all and thus are vulnerable to undetected loss. See Appendix C for a brief discussion of the differences.

**Figure** 3 illustrates the effectiveness of simple annual auditing across a very wide range of sector failure rate over a long period.

**With regular auditing, only a few copies are required to minimize losses over a wide range.**
**Failure to audit the collection is worse than keeping**
**only a small number of audited copies**

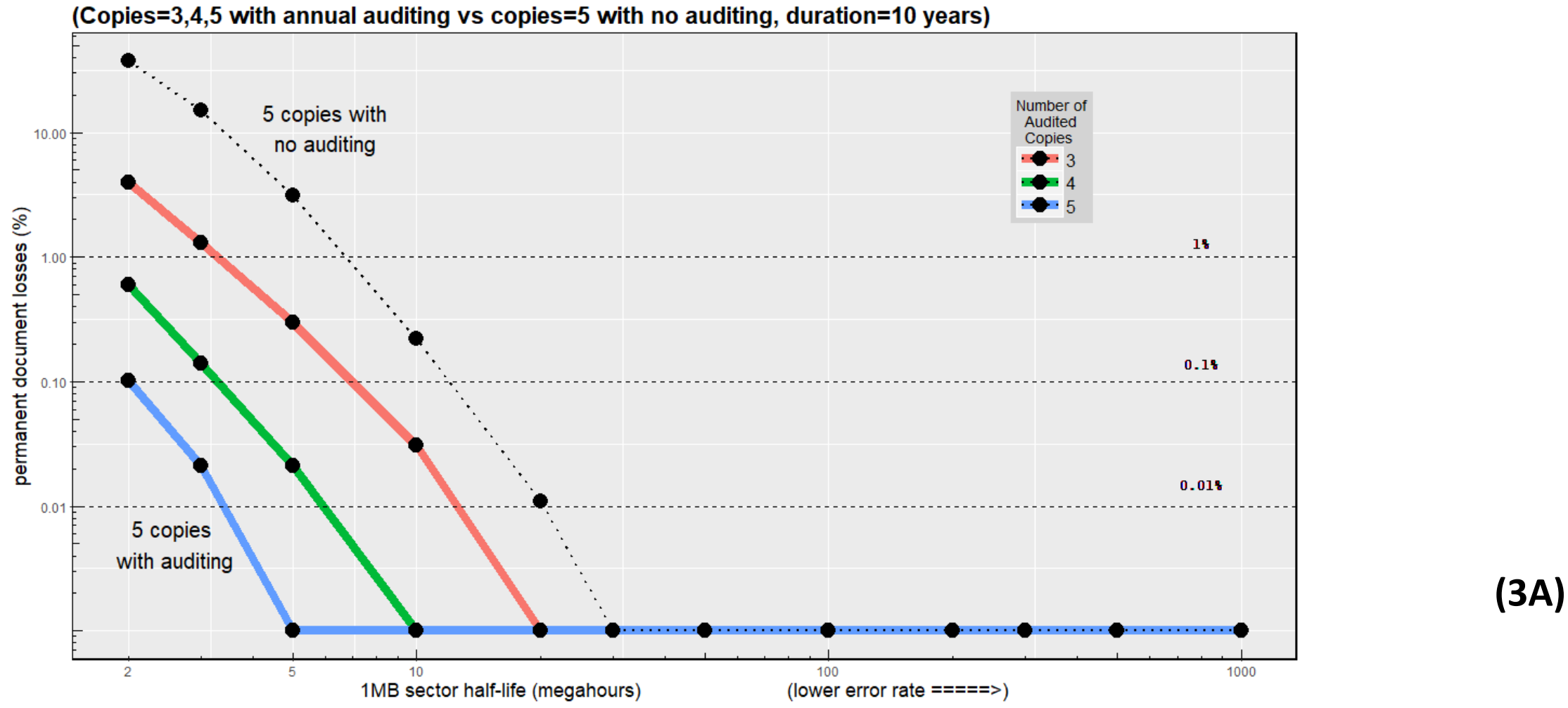

**(3A)**

**Can a collection survive long-term with enough copies and annual auditing?**
**100-year document survival based on number of copies and disk error rates**

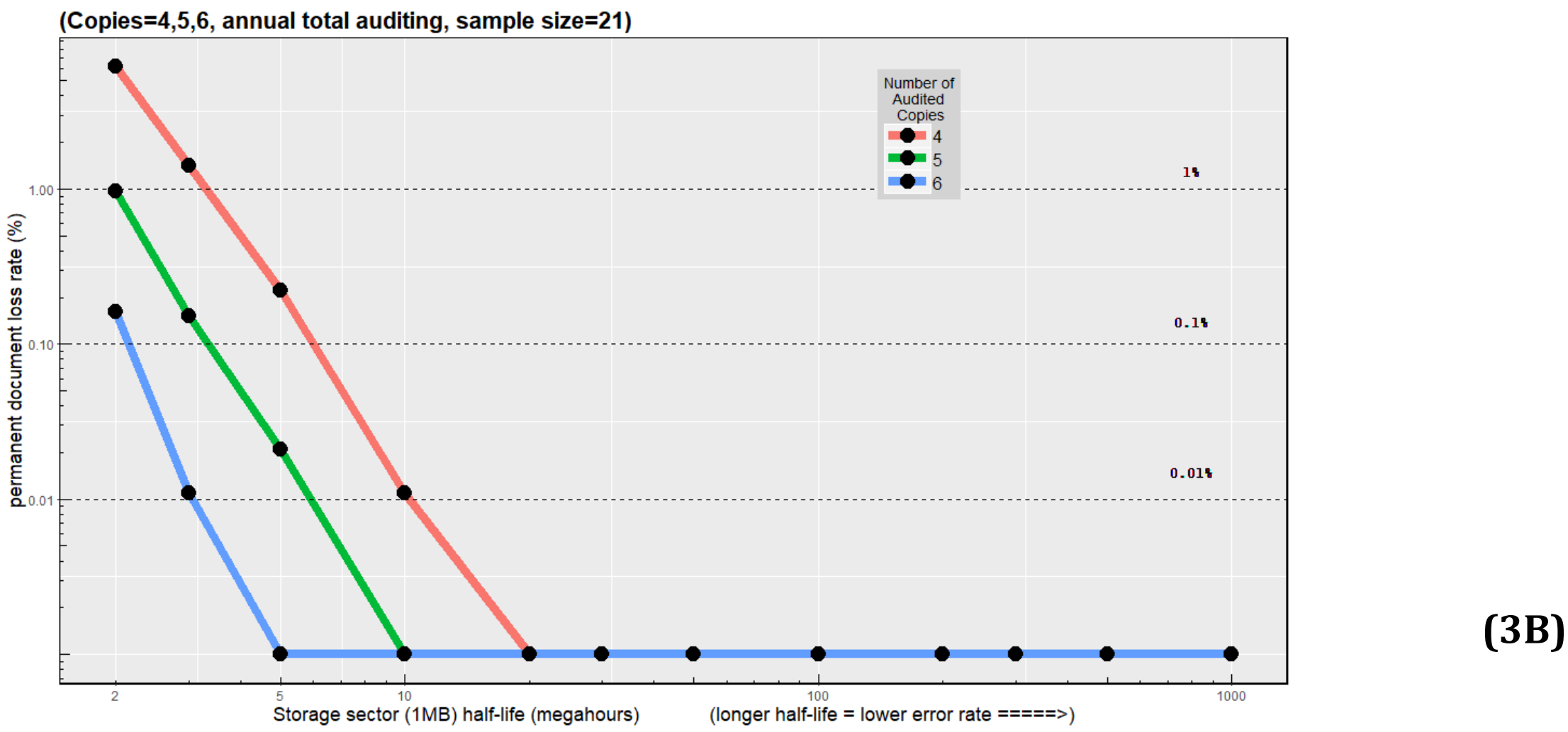

**(3B)**

**Figure 3: Simple annual auditing and repair of a collection greatly reduces document losses. (3A) Note that just three copies with annual auditing is more robust than five unaudited copies. (3B) Even over very long periods of time, e.g., 100 years, simple annual auditing can protect a collection with a modest number of copies. Over the plausible range of sector lifetimes, five copies with auditing suffer minimal or zero losses.**

The pattern illustrated by the figure (and confirmed through additional sensitivity analysis) demonstrates that a strategy of deploying five independent replicas and simple annual auditing is sufficient to maintain the integrity of collections for over a century across any practical variation in storage quality conditions. In the full paper, we extend these results to include robustness to higher-level correlated threats, and to analyze more sophisticated and efficient preservation strategy.

**Constructing a Preservation Strategy that is Robust to Institutional Risks**

**Strategies against limited server lifetimes or shocks**

Cloud-based storage services provide very reliable storage, largely through the use of error-correcting storage methods, such as RAID, erasure coding, and high-count replication. These error-correcting techniques make the failure rate for disk sectors largely irrelevant. Such storage, properly maintained, is effectively "immortal" and will suffer no significant losses over long periods.

Let us shift our analysis from the reliability of storage media to the reliability of storage services. We will find that similar analysis and similar storage techniques can be used to protect documents and entire collections.

Storage services themselves are not immortal. Services as corporations may fail over time; they may merge and thus lose their independence; they may be subject to physical trauma through natural events or political or economic disturbances.[6] And access to services may be blocked by government action, by cyber attack, or by administrative error (such as a problem with service billing). In any case, it is important to note that such failures are generally silent: the client is not actively informed of the failure. The client will notice the failure only when trying to access the stored documents.

There are many ways that a server can fail and render a copy of the collection inaccessible. Table nnn lists a number of conditions that can cause one or more servers to fail. Let's consider just two causes of server failure: first, that servers as corporations have finite lifetimes; and second, that exogenous physical events or government actions may make it impossible for a server to continue to function. We note that a shock that raises the rate of failure of a single server is equivalent to a server with reduced life expectancy.

To protect a document collection, a client library must actively test, and if necessary repair, the integrity of document collections stored remotely on such services. Server failures, regardless of the cause, are almost always silent to their clients: earthquakes, floods, wars, mergers, bankruptcies, and government censorship actions do not give notice to the parties affected. A client must examine all servers on a regular basis to verify the presence of the document collection stored there. The client must actively audit storage services to verify that they are still alive and still have the collections. Due to the assumed high reliability of storage within a server, we assume that a server contains all documents or none. If a service is still alive and contains any documents, then the service is still available as one of the client's replication instances. If a service is found to be inactive, then, to maintain the target number of replications, the client must find a new service and populate that with the collections.

This process is clearly analogous to the auditing of individual documents to repair documents corrupted by sector failures, but in this case, it is merely the presence of the server that is to be tested. We find also that the auditing process can be much more efficient than with sectors: to verify that the server is alive, only a few documents need to be retrieved. Either the server is dead and contains no documents, or it is alive and contains all documents.

[6] Reviews of firm mortality suggest that the half-life of firms is no more than a decade, and likely shorter in the technology sector and during recessions. (Morris 2010; Daepp 2015) Recessions are frequent relative to this lifespace --see (NBER 2019) for data on US recession frequency.

When a client detects a server failure of this sort, the redundancy of the collection storage is (temporarily) reduced. The client must find another server to hold a copy of the collection and then transmit the entire collection to that server. Only then is the collection fully replicated with the desired number of copies.

The simulation framework demonstrates that server lifetime is an important risk factor. The replication strategy previously developed (shown in Figure 3), although robust to sector failures and glitches, will fail (that is, result in collection losses) if the server failure rate is high. For that simple strategy -- five copies audited annually -- to preserve a collection over long periods of thirty to one hundred years would require servers to have half-lives of at least eight to ten years. To require all servers to have such long life expectancies is optimistic. However, on the bright side, collection loss can be prevented even with relatively short-lived servers by substantially increasing the auditing frequency and adding additional replicas.

### Strategies against correlated server failures

The strategy above is robust to server failure -- when servers fail independently. However, server failures may be correlated in two ways. We model these correlations formally, through introducing "shock" events of varying types, severity, and frequency.

We use two statistical approaches to model shocks. First, we model shock that *decrease the expected lifespan of some or all servers.* Second, we model shocks that *cause immediate failure of multiple servers simultaneously.* Our simulations included a range of "shock" conditions that contribute to server failures; modeling such threats as recession, targeted censorship, administrative error.

For the most part, our results show that the most important factors are the frequency of shocks and their "span," that is, the number of (correlated) servers affected by a shock. Overall, shocks that raise the rate of single server failure have an impact equivalent to reducing the expected lifespan of all servers. That is, whether the cause of a server's premature failure is exogenous, through economic or political pressure, or endogenous, due to financial instability, the result is the same: a single server fails at some random time with some frequency independent of other servers. This failure reduces the redundancy of the storage of the collection and therefore increases the risk of collection loss.

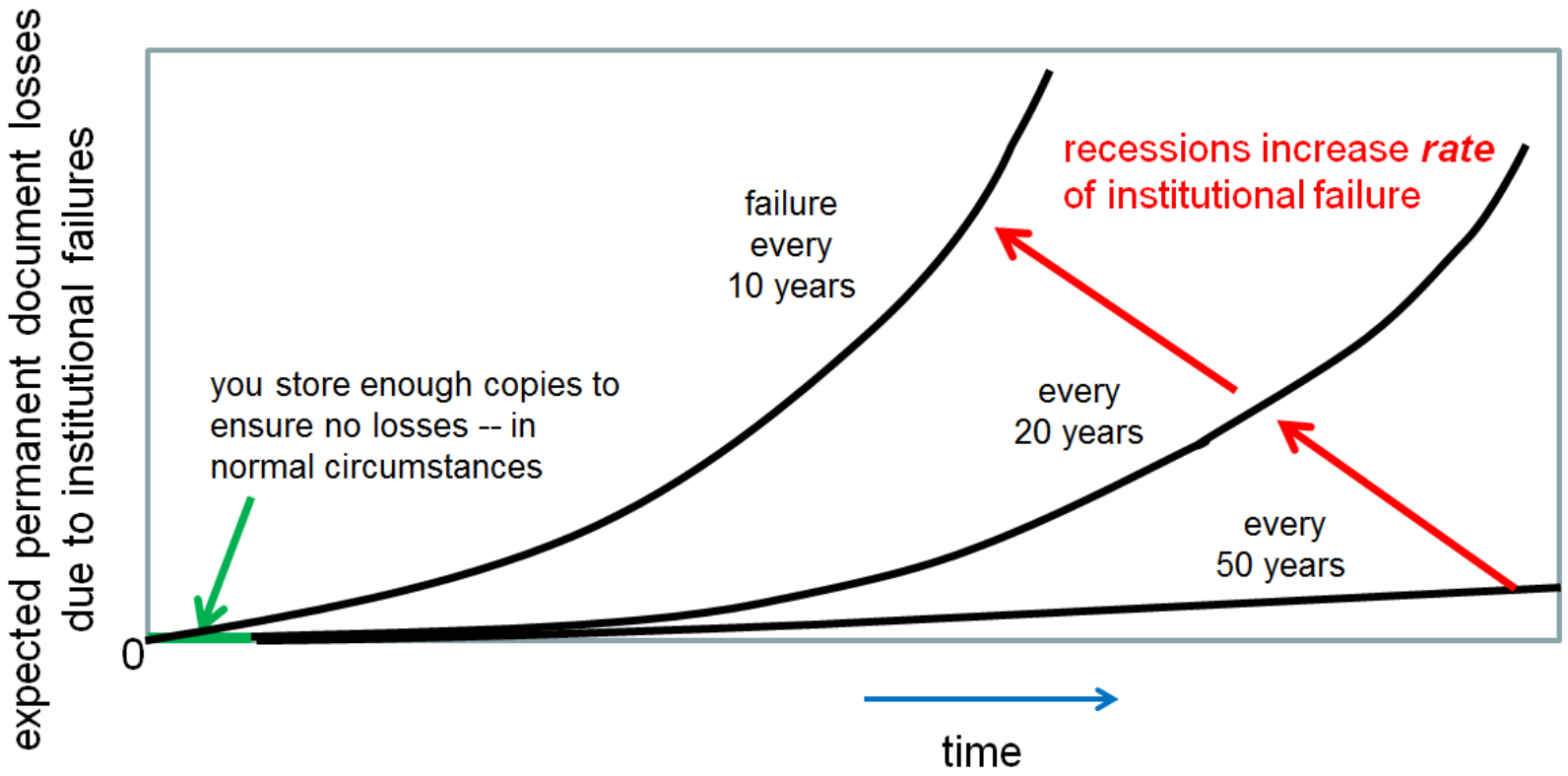

**Figure 5: The effect of economic pressures on institutional failures. For any chosen level of redundancy that protects the collection in normal circumstances, economic recessions will increase the rate at which the servers that keep copies fail. Severe recessions will cause institutional failures sooner than mild recessions, but still, over time, some institutions will fail. Without a strategy to detect and repair these failures, the redundancy of collection storage is reduced and the likelihood of collection loss is increased.**

In contrast, shocks that cause multiple simultaneous failures can dramatically increase the likelihood of collection loss. In our simulations, we have found that the major protective factor against correlated failures is the speed of detecting a dead server. The client should test all the servers for responsiveness very frequently, such as quarterly or monthly; in very severe cases, testing should be done even faster, e.g., weekly .

Again, it is important to note here that it is not necessary to test the entire collection, or even a large segment of it, during each audit. Since a server either contains all documents or contains none, the client need only probe a few documents during each audit to determine the health of the server. This dramatically reduces the bandwidth requirements and increases the speed with which a client can test all its servers.

Thus a strategy of having X copies audited every Y period can be robust even in the presence of frequent widespread shocks: such a strategy is likely to withstand both major recessions and minor wars. As **Figures 6 and 7** below illustrate, protecting a collection under conditions of frequent large shocks, economic or political, may require increasing the redundancy of storage to seven or eight copies, and increasing the frequency of auditing of servers from quarterly or monthly to weekly.

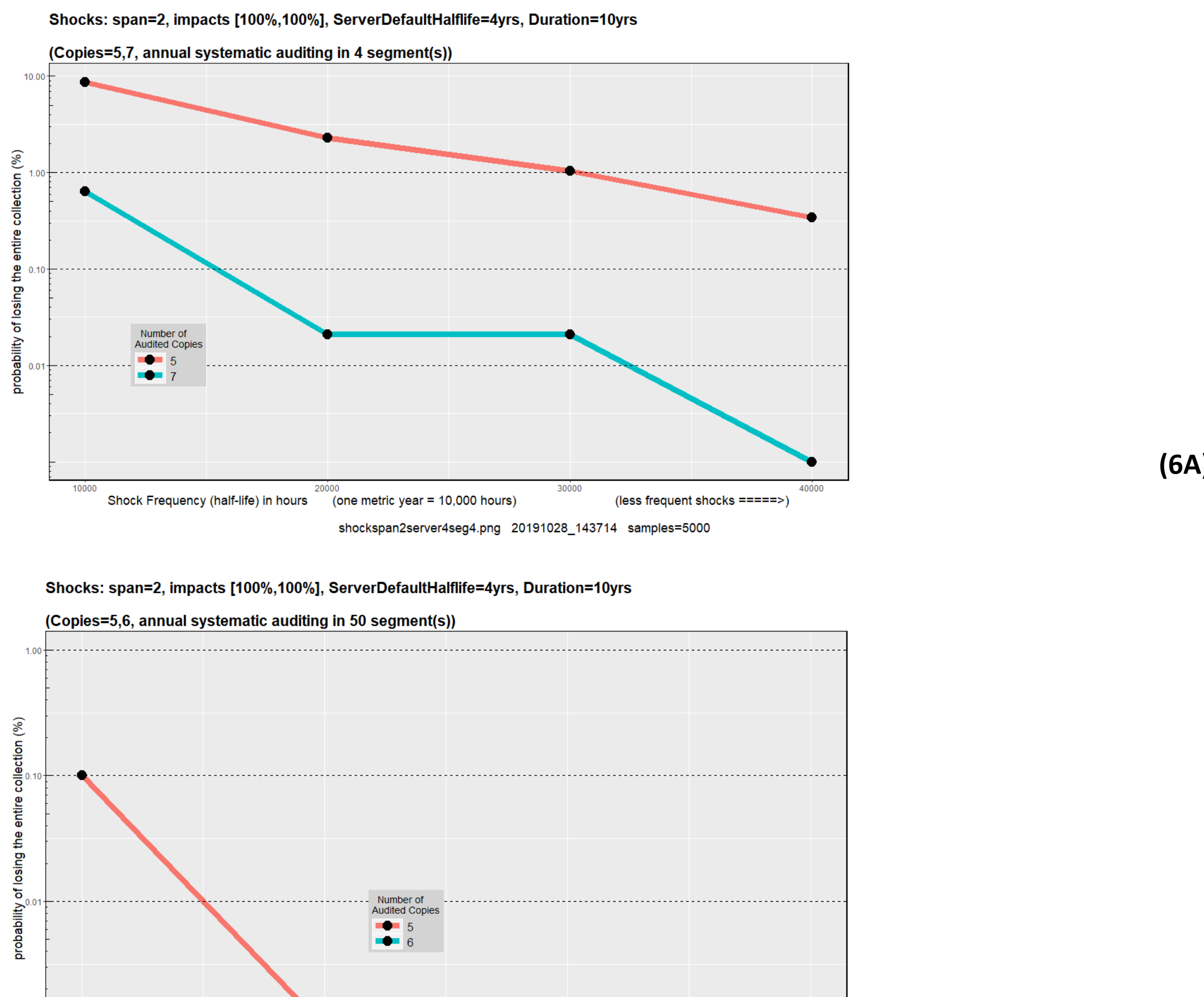

**Figure 6: The impact of moderately severe correlated failures of servers over ten years. In these cases, shocks that cause two servers to fail immediately occur randomly with half-lives shown on the X axis. (6A) Five copies of a collection with quarterly auditing is not sufficient to protect against a high probability of total loss. Increasing the redundancy to seven copies also does not protect the collection sufficiently. (6B) Accelerating the rate of auditing, e.g., to weekly segments, and possibly increasing the redundancy level slightly, can improve the survival of the collection even under these severe conditions. Recall that auditing to determine the presence of a server requires interrogating only a few documents on each server.**

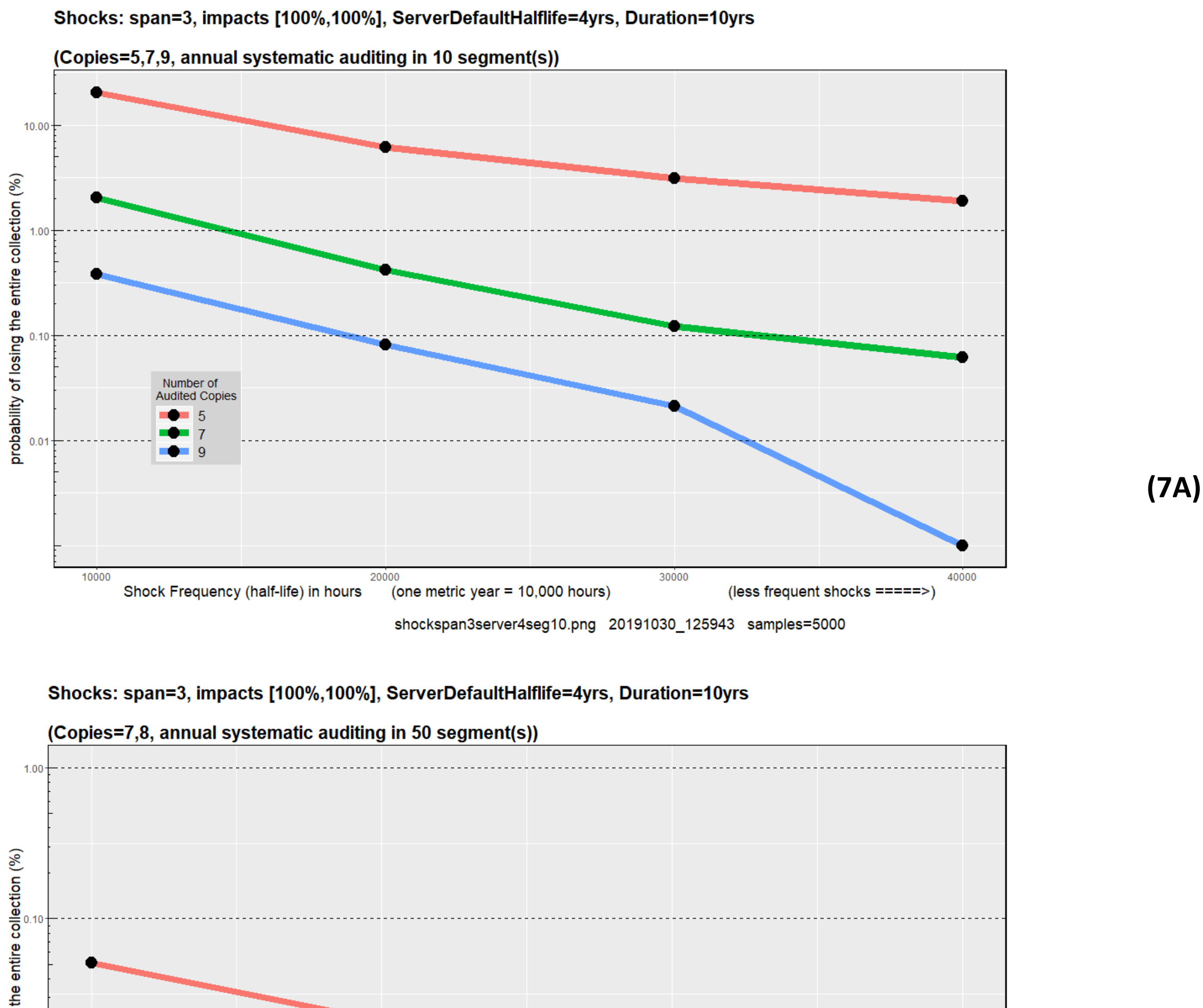

**Figure 7: The impact of very severe and frequent correlated failures of servers over ten years. In these cases, shocks that cause three servers to fail immediately occur randomly with half-lives shown on the X axis. Much more frequent auditing of more copies is required to protect a collection. (7A) For shocks where three servers are removed from service, even frequent monthly auditing with many more copies, up to nine, cannot protect the collection from total loss. (7B) Accelerating the auditing to weekly and increasing the redundancy to eight copies may suffice to protect a collection from total loss even under these extremely harsh conditions.**

We note that a combined strategy might be a good choice: Audit the collection on an annual cycle in weekly segments; every week choose two percent of the collection, selected *without replacement*, to be audited. Replace any servers found to be not functioning with new servers, and repopulate the new servers with the entire collection. Using this strategy, redundancy of five copies should protect the collection in all but the most dire circumstances. Increasing the redundancy to seven or eight copies should protect the collection even in very severe upheavals.

**Corollaries - Applications of the Multi-Level Model to Other Failures**

In this section we show how the existing simulation results can be applied to modeling additional threats, including attacks against the auditing system; encryption-key loss and other related correlated failures; and fragility induced by document compression.

**Augmenting the Core Replication Strategy with Distributed Auditing**

In the model above auditing always returns the correct results, when invoked. In practice, the auditing system itself may fail either due to a fault in the auditing software, or to a malicious attack against the auditing system.

Risks due to unintentional faults in the auditing system may be mitigated by using multiple independent software implementations to perform the proof-of-retrieval. For example, in an auditing system that involves retrieval of the object content, followed by computation of a cryptographic hash on the contents; computation of this hash could be replicated using multiple independent implementations of the hashing algorithm. More generally, to prevent loss due to well-resourced attacks against the auditing system itself, a secure auditing system that engages multiple auditors should be used [see Jin et al 2019 for a review]. Further, since in the worst case, a well-resourced adversary could subvert one or more of the auditors, at least $3s+1$ servers will be needed to provide *byzantine fault tolerance* -- where $s$ is the maximum number of subverted nodes.

Thus in a world without any types of server failures (besides malicious attack), deploying seven servers would be sufficient to protect against subversion of two. Distributed preservation systems such as LOCKSS, have each server play the role of both a content replica, and an independent auditor. When using distributed preservation, or in other systems in which auditing servers can fail at random, additional replicas will be needed to ensure that sufficient *uncorrupted* replicas remain. For example, if there are two potentially subverted servers, and external shocks can destroy three servers simultaneously at random, then post-shock, two out of four of the remaining servers could be corrupted. Thus with a shock-span of three, ten replicas are necessary to prevent against orchestrated attacks by two subverted noded on the auditing system that are launched following a failure. Eight copies would be sufficient, if the adversary controls at most one auditing node.

**Applying Analysis to Replication of Encryption Keys**

We identified loss of encryption keys (or more generally of shared secrets -- where at least n are needed to reconstruct the key) as a threat to content above. In this section we discuss strategies for evaluating the impact of encryption key loss, and mitigating these risks. As it turns out, it is not necessary to add encryption directly to the discrete simulation model described above -- we can evaluate this risk using the existing model.

Encrypting a collection of content creates four additional threats of loss. First, and most important, losing all copies of the encryption keys for the collection effectively results in a loss of all replicas of the collection -- while the bits comprising such collections may continue to exist, they are rendered meaningless. Second, if the knowledge of the encryption algorithm is lost, the collection is likewise destroyed. Third, encryption may make documents more fragile -- a single block loss may destroy the entire document rather than a portion. Finally compression, which decreases risk of loss (see below), is ineffective on encrypted files.

The last two threats are minor: In most cases compression can be applied effectively prior to encryption, and the storage savings can then be used to add an additional replication copy which more than offsets the fragility increase from both the compression and encryption. (See *modeling compression risks,* below, for more details) The second threat (algorithm loss) can lso be effectively mitigated by selecting a well-known standard encryption algorithms -- standard algorithms are widely documented, and independently replicated. Thus, we focus on the threats from loss of encryption keys.

Loss of encryption key may be modeled by treating the keys themselves as a small, separate collection of documents. As we have shown above -- mitigating risks of loss to a 'key-collection' requires replication, auditing, and repair. Because the size of the collection is small (encryption keys are very small relative to the content they protect), risks to the collection of keys will be dominated by 'shocks' that disrupt organizations and affect multiple replicas: For example -- wars, economic recessions, and government actions may lead to organizational failures

In summary modeling encryption loss involves extending the semantics of 'collection' and then reinterpreting the simulation results for server-failure in this new context: In this case, the risks from loss of encryption keys is equivalent to the risk of loss of an unencrypted collection that is subject to correlated shocks. Thus, the simulation results summarized above (especially in Figures 6 and 7) can be used to predict encryption key failures under different conditions. There are two practical differences from typical collections to consider in this case. For information security best practice, an encryption keys should be kept in a separate administrative domains than the content they encrypt -- so some care must be used in assigning replicates to specific servers. And in practices the size of encryption is very small relative to collection of content -- this creates the opportunity for more frequent auditing.

Thus based on the simulation results already discussed above we recommend maintaining at least four copies of the encryption key-collection and auditing the integrity of the key-collection (e.g. using a challenge mechanism) frequently -- at least monthly. We also recommend that a separate set of 'key-servers' be used to replicate the collection of encryption keys where possible -- or at minimum that no server contain replicates of its own key. [7]

### Modeling Risks of File-Format Obsolescence

At first blush, file-format obsolescence appears to present a radically different form of risk that the bit-loss modeled. After all, a collection may be lost to format-obsolescence even if all of the replicas are intact -- instead, file format obsolescence occurs when no available software is capable of successfully 'reading' files in that format.

Bit-level auditing does nothing to prevent this. It is possible to model auditing for a file-format obsolescence in a way that parallels detecting server-level failure, and thus use server- and shock-analysis to estimate risks of loss, and to develop mitigation strategies. This is accomplished by using the same simulations approach as above, but reinterpreting the semantics (i.e. relabeling) of the model to describe testing corpuses of documents with indepenent readers. In short, if reader can no longer be executed to read a corpus in a formt, it has “failed”, and if all readers fail, the format is determined to be obsolete. More specifically, in the relabeled model:

- A 'format-collection' is redefined as a collection of documents stored in a designated format. This corpus will be used to check for formal loss.

[7] For the purpose of this analysis, we assume any single key is sufficient to decrypt content. In cases where external parties are not fully trusted, but can be prevented from colluding en masse, secret sharing schemes algorithms can be used to, in effect, to split keys such that any *m* keys together will decrypt content. (see Beimal 2010, for a survey of methods). In this case an additional *m-1* key replicates should be provisioned.

- As in the model of encryption key loss above, the collections of test documents are relatively very small -- and thus sector-level error is ignorable.
- A 'format-server' represents a server operating an independent implementation of software that can (a) accept a collection as input (b) validate that has 'read' the collection files, in the designated format.
- Auditing a format server consists of executing the corresponding reader against the test corpus.
- A format server failure occurs when it is discovered on auditing that a particular collection cannot be read. In other words, the format is no longer interpretable by that reader implementation.
- When all servers fail the format becomes obsolete (by definition).

Using this remapping we can now interpret the proability of collection loss due to shocks as the probability of format obsolescence -- assuming a probability distribution for correlated failures over independent readers. However, unlike storage systems, there is substantially less evidence to assess the distribution of failures of software readers -- so we proceed conservatively.

Since it is likely that software failures follow a modified bathtub curve an assumption of an exponential distribution of reader failure would yield overly optimistic predictions for loss format obsolesence. Notwithstanding distributional assumptions, by conditioning on using format readers that are established (past their infant mortality period), modeling with a conservative (lower than expected) expected lifetime, and modeling shocks that cause multiple readers to fail simultaneously, we can generate strategies that are robust to format failure.

Since software reader failure is driven primarily by changes in the operating environment, we anchor our half life estimates to the support lifetime of operating systems version. In particular, a plausible but very conservative assumption is that half of software readers fail to run without modification after a major operating system revision. Over its thirty year history, the Windows operating systems has undergone a major update every six years on average, and the supported lifespan of each revision has averaged twelve years [Wikipedia Contributors 2019]. Thus, based on the server failure analysis shown in figure 4b, using five independent readers to test format readability at each major software release, accompanied by migration of formats, when five functioning independent format readers can no longer be identified, should be sufficient to protect against format failure indefinitely. Further, as the analysis above shows, the risk of failure is strongly dependent on the frequency of auditing: Thus a strategy of format assessment using three independent readers should be successful indefinitely, if a more sophisticated timing of audits is planned -- e.g. verifying readability annually, and in advance of planned operating system updates and of support end-of-life. dates.

### Modeling Robust Documents and Compression Risks

Documents stored on digital media are fragile; storage errors corrupt the content of a document. How much of a document is corrupted depends largely on the data format of the document. Even small errors in highly compressed or encrypted documents may render part or all of the document unusable.

For documents that might not be fatally corrupted by a single sector error, lossless compression of the document involves a clear trade-off. A smaller document is a smaller target for a randomly occurring error, but a highly compressed document is more fragile. A small error in an audio or video file, or an uncompressed text file, might not be fatal to the document, but a highly compressed text document (or an encrypted document) might be lost.

In these simulations, for simplicity we have modeled documents as very fragile: one sector error causes the document to be judged as lost. As it turns out, straightforward closed-form transformations of the previous simulation results can be used to model the effects of lossless file-level compression and document fragility on overall collection loss.

In this model, at least these two considerations should be included in the decision to compress documents.

- Smaller is safer. A smaller document presents a smaller target for random errors. If a document is compressed, say, by 90%, that is, to 10% of its original size, then a random error is only one-tenth as likely to strike that document. When placed on a storage medium of any given quality level, that smaller, compressed document is likely to persist without error ten times longer than the uncompressed version.
- Smaller is less expensive. A stored collection incurs costs for both storage of the document images and the bandwidth used in auditing and repair. Smaller documents consume less space and less bandwidth and therefore cost less to maintain. On a given budget, a compressed collection can be replicated into more copies and audited more frequently. Both the increased copy count and more frequent auditing contribute directly to reducing or eliminating permanent losses in the collection.

The linear increase in document losses based on size is to be expected from straightforward Poisson calculations. In addition, we ran simulations over a wide variety of conditions to verify that this linear relationship holds for multiple copies of collection documents, various auditing strategies, and a very wide range of storage quality (error rates, sector lifetimes).

Compression offers another major advantage: potentially higher redundancy. If compression reduces a document's size by, say, 50%, then a client can store two copies of the document for the same cost in storage. That extra copy provides higher redundancy and thus greater resistance to document loss. On a fixed budget, a client can store additional copies of documents depending on how effective the compression algorithm is. High compression permits more copies to be replicated to offset any increased fragility of a compressed document. Text and image compression are particularly effective in this regard.

A disadvantage of compression is that it may make documents that were partially repairable more fragile. We use a simple model to explore the effects of fragility. In a given collection, the fragility of each document is represented by a number, *F,* that indicates the proportion of damage done by each single sector loss. For example, an *F* of 1 indicates that single-sector loss causes the entire document to be lost; an F of 2 denotes that a single-sector loss reduces the value of the document by half, etc.

For the purpose of loss prediction, a collection of robust (not fragile) documents can be modeled as a more numerous collection of smaller, fragile documents. More precisely, the expected proportion of losses due to block failure for a collection *C* of *N* documents, each of size *S* and fragility *F* is identical to the expected proportion of losses in a collection C' comprising (N*F) documents, each of fragility 1, and size S/F.

Thus, lossless document-level compression affects the likelihood of collection loss through four distinct mechanisms. First, compression directly reduces document size (modeled as a 'compression ratio' of 1/*C*), which acts directly to reduce expected loss. Second, compression can increase the fragility of the compressed document *F'*, where $F \geq F' \geq 1$. Third, compression reduces the total size storage size needed for the collection by -- enabling more replicas to be created.[8] Finally, compression permits more aggressive document auditing to protect a collection without increasing.

[8] In addition compression may introduce a risk of format obsolescence. Fortunately, there are effective file-level compression algorithms that are very well standardized and documented. Further, the risks from format obsolescence may be managed effectively through auditing format-readers, as described above.

costs for bandwidth and server egress. The same number of smaller, compressed documents can be retrieved more quickly without increasing bandwidth and consume less bandwidth and less egress charge from the storage vendors. Auditing of the collection can be done more frequently on the same budget, which improves document survival rates.

The effects of additional replicas can be understood by considering the “repair” of collections through the replacement of documents. In our model, collections are "repaired" by having missing documents on their servers replaced from other copies; that is, the "repairability" of a collection depends on the presence of one or more valid replacement copies stored elsewhere. If compression permits an additional copy or copies of a document to be stored, then there will be more copies from which a replacement can be effected when one copy fails and a collection needs to be "repaired." For example, if five copies of a collection are to be stored on servers, then a mere 20% reduction in size due to compression would permit one additional copy to be stored and maintained within the same budget. That additional redundancy, six copies instead of five, would make the document more resistant to failure.

These effects can affect loss in a different direction -- when does compression reduce loss overall? The simplest case is of completely fragile documents (F=1) -- in this case, F’=F and any amount of compression is beneficial due simply to reduced “target area.” More generally, compression will reduce loss whenever $C*F' \geq F$. Further, the simulation results above demonstrate that even where sector error is the dominant threat to collections (which is not generally the case) the benefits of increasing the number of copies grows substantially up to at least six (6) replicas. An implication is that compression with C of 1.2 (reduction to roughly 83% of the original size) or better is usually justified even when it substantially reduces document fragility over ( $F >> C*F'$ ). Since, in practice, modern compression algorithms yield a *C* typically in the range of 2-6 on benchmark corpora [Mahohen 2005] for large collections; and methods to reliably repair damaged documents are scant and hard to verify, compression reduces expected collection loss in all but a few extreme cases.

In summary, we consider lossless compression to be benign for a variety of reasons.

## Discussion

The results above highlight a number of robust and broadly applicable operational preservation policies: for example, these results demonstrate that the commonly used strategy of sample-based auditing is ineffective; and that the risks of compression-related fragility noted by the preservation community, are typically more than offset by reductions in the efficiency of replication and auditing. More generally, we identify simple preservation strategies involving diversification, 5-7 replicates, and auditing partitioned weekly across every year, that are robust both to variations in storage quality and conditions; and robust to correlated organizational threats, including global recessions and regional wars.

This analysis demonstrates that the most critical source of risk to collections is shared vulnerabilities across services that can result in multiple simultaneous failures. Curators who choose services for replication need to be wary of characteristics that result in shared vulnerabilities. These include the geographic location of server infrastructure; legal jurisdictions to which the server is subject; and shared technical dependencies across servers. It is particularly important that service providers disclose in a verifiable way the extent to which they rely on other third-party storage vendors to host content stored with them.

Further, these results underscore the need to increase the efficiency of external auditing since auditing frequency is critical to robustness. Currently, a bottleneck to auditing is the need to transfer the content of a document to be audited from a server to a trusted source. If storage services provided an API for trustworthy verification of a document directly, costs and time would be substantially reduced. For instance, if a storage service offered the ability to compute, on-demand, a combination cryptographic hash and nonce for a selected portion of a document, reliable external auditing could be completed with greatly reduced network bandwidth. (See Lin, et *al*. 2019 for a survey of cryptographic auditing approaches to cloud replicas.) Stakeholders in both the storage and preservation sectors would benefit from developing standards APIs for cryptographic verification of stored content -- which increase both the trustworthiness of cloud services and the use of external auditing.

Moreover, the framework we present offers flexibility for the modeling of a wide range of preservation policies and threats. Since this framework is open source and easily deployed in a cloud computing environment, it can be used to produce analyses based on independent estimates of scenario-specific costs, reliability, and risks. We invite the community to probe our results by calibrating risk profiles based on their own context and using the system to estimate the costs and losses of their preferred preservation strategies.

# Acknowledgments

The authors describe contributions to this article using a standard taxonomy. [Allen, et al. 2014] MA provided the core formulation of the research goals and aims, and methodology. RL led the data collection, software design, and data analysis. MA and RL shared the writing of the original manuscript. All authors contributed through commentary, review, editing, and revision.

# References

Allen L. , Jo Scott, Amy Brand, Marjorie Hlava & Micah Altman, (2014). Publishing: Credit Where Credit Is Due, 508 *Nature* 312. <https://doi.org/10.1038/508312a>

Altman, *et* al., National Agenda for Digital Stewardship, 2015 *National Digital Stewardship Alliance* (2015) <https://osf.io/23vph/>

Altman, *et* al., (2020). 2020 National Agenda for Digital Stewardship, *National Digital Stewardship Alliance*. <https://osf.io/bcetd/>

Baker, M., Shah, M., Rosenthal, D. S., Roussopoulos, M., Maniatis, P., Giuli, T. J., & Bungale, P. (2006). A fresh look at the reliability of long-term digital storage. In *ACM SIGOPS Operating Systems Review* (Vol. 40, No. 4, pp. 221-234). ACM. <https://dl.acm.org/doi/10.1145/1218063.1217957>

Beimel, A. (2011). Secret-sharing schemes: a survey. In *International conference on coding and cryptology* (pp. 11-46). Springer, Berlin, Heidelberg. <https://doi.org/10.1007/978-3-642-20901-7_2>

Daepp, Madeleine IG, *et* al. (2015). "The mortality of companies." *Journal of The Royal Society Interface* 12.106 <https://doi.org/10.1098/rsif.2015.0120>

Gallinger, M., Bailey, J., Cariani, K., Owens, T., and Altman, M., (2017). Trends in Digital Preservation Capacity and Practice: Results from the 2nd Bi-annual National Digital Stewardship Alliance Storage Survey*, D-Lib Magazine:* 7/8 <https://doi.org/10.1045/july2017-gallinger>

Lebrecht, A. S., Dingle, N. J., & Knottenbelt, W. J. (2011). Analytical and simulation modeling of zoned raid systems. *The Computer Journal*, 54(5), 691-707. <https://doi.org/10.1093/comjnl/bxq053>

Li, Y., Miller, E. L., & Long, D. D. (2012). Understanding data survivability in archival storage systems. In *Proceedings of the 5th Annual International Systems and Storage Conference* (p. 16). ACM. <https://doi.org/10.1145/2367589.2367605>

Lin, Y, Li, J, Jia, X, Ren, K. (2019). Multiple-replica integrity auditing schemes for cloud data storage. Concurrency Computational Practice & Experience: 33(7).<https://doi.org/10.1002/cpe.5356>

Mahoney, Matthew V. "Adaptive weighing of context models for lossless data compression." (2005). *Florida Institute of Technology CS Dept, Technical Report CS-2005-16*. <http://hdl.handle.net/11141/154>

Mellor, C., (2018). Mmm, yes. 11-nines data durability? Mmmm, that sounds good. Except it's virtually meaningless, *The Register*, <https://web.archive.org/web/20210227122023/https://www.theregister.com/2018/07/19/data_durability_statements/>

Morris, James R. (2009). "Life and death of businesses: A review of research on firm mortality." *Journal of Business Valuation and Economic Loss Analysis:* 4(1). <https://doi.org/10.2202/1932-9156.1050>

NBER (National Bureau of Economic Research). (2019) "US business cycle expansions and contractions." <http://www.nber.org/cycles/cyclesmain.html>

Pinheiro, E., W.D. Weber, L.A. Barroso, (2007). Failure trends in a large disk drive population. In *Proceedings of the 5th USENIX Conference on File and Storage Technologies* (FAST '07). <https://www.usenix.org/legacy/events/fast07/tech/full_papers/pinheiro/pinheiro_old.pdf>

Rosenthal, David S.H.; T. Robertson, T. Lipkis, V. Reich, and S. Morabito, (2005). "Requirements for Digital Preservation Systems", *D-Lib Magazine* 11 (11). <https://doi.org/10.48550/arXiv.cs/0509018>

Rosenthal, David S.H. (2010). "Bit preservation: A solved problem?." *International Journal of Digital Curation* 5(1): 134-148. <https://doi.org/10.2218/ijdc.v5i1.148>

Wikipedia contributors, "Microsoft Windows," (2019) Wikipedia, The Free Encyclopedia, (Retrieved on December 15, 2019, from: <https://en.wikipedia.org/w/index.php?title=Microsoft_Windows&oldid=929717556>

Xin, Qin, Thomas JE Schwarz, and Ethan L. Miller. (2015) "Disk infant mortality in large storage systems." *13th IEEE International Symposium on Modeling, Analysis, and Simulation of Computer and Telecommunication Systems*. IEEE. <https://doi.org/10.1109/MASCOTS.2005.27>

# Appendix A

The rate at which a stored document deteriorates depends on its size as well as on storage-quality and other factors. We selected a nominal document size for the simulations to avoid redundancy in the choices of parameters. A corollary of the model structure is that the rate of accumulation of errors, in theory, depends linearly on the size of the document. We checked this prediction with simulations using documents of differing sizes and error rates. The results are shown below.

Note that as the size of a document increases from 5 megabytes (MB) to 5000 MB, if the rate of error arrivals decreases inversely, then the total document losses remain constant. For example, a collection of 5 MB documents stored with a sector lifetime of 5 megahours has the same loss rate as a collection of 5000 MB documents with a sector lifetime of 5000 megahours. This linear relationship holds across at least these several orders of magnitude that encompass plausible document sizes and storage qualities.

The spectrum of sizes ranges from 5 MB, representing a PDF document or small still image, to 5000 MB, representing a large video file.

**Document Size vs Error Rate Comparisons**

**Comparisons: larger docs and corresponding increase in sector lifetimes yield the same loss rates**

| Document size | Sector Lifetime (half-life) | Percent of Collection Lost |
|---|---|---|
| 5 | 2 | 15.9 |
| 50 | 20 | 15.9 |
| 500 | 200 | 15.9 |
| 5000 | 2000 | 15.9 |
| 5 | 3 | 10.9 |
| 50 | 30 | 10.9 |
| 500 | 300 | 10.9 |
| 5000 | 3000 | 11 |
| 5 | 5 | 6.6 |
| 50 | 50 | 6.7 |
| 500 | 500 | 6.7 |
| 5000 | 5000 | 6.7 |
| 5 | 10 | 3.3 |
| 50 | 100 | 3.4 |
| 500 | 1000 | 3.4 |
| 5000 | 10000 | 3.4 |

# Appendix B

To confirm the calibration of the simulation methods, we compared the simulation results with the theoretical Poisson calculations, using a very modest number of repetitions. The simulation results compare favorably with the calculated values. Over three orders of magnitude, the simulations match expectations very closely; larger simulations would be expected to match an even broader range of parameters.

**Poisson Failure Rates based on HALF-LIVES**

Document hits over duration of run

But since a document is much larger than a single unit,
the effective lifetime is reduced, because the target area expanded,
by that factor. Which is fifty in this case.

| param --> | half-life | lifetime = 1/ln(2) | t | Ntargets | target size | effective lifetime | Pr(0 hits) on a target | theoretical losses | Simulation results n = 101 reps midmeans V V V | err |
|---|---|---|---|---|---|---|---|---|---|---|
| units--> | khours | khours | khours | k | MB | khours | | Ntargets | copies=1 | pct |
| | 1,000 | 1,443 | 100 | 10 | 50 | 29 | 0.0313 | 9687.5 | n/a | n/a |
| | 2,000 | 2,885 | 100 | 10 | 50 | 58 | 0.1768 | 8232.2 | 8227.5 | -0.1% |
| | 3,000 | 4,328 | 100 | 10 | 50 | 87 | 0.3150 | 6850.2 | 6849.6 | 0.0% |
| | 5,000 | 7,213 | 100 | 10 | 50 | 144 | 0.5000 | 5000.0 | 5004.0 | 0.1% |
| | 10,000 | 14,427 | 100 | 10 | 50 | 289 | 0.7071 | 2928.9 | 2929.7 | 0.0% |
| | 20,000 | 28,854 | 100 | 10 | 50 | 577 | 0.8409 | 1591.0 | 1589.3 | -0.1% |
| | 30,000 | 43,281 | 100 | 10 | 50 | 866 | 0.8909 | 1091.0 | 1088.5 | -0.2% |
| | 50,000 | 72,135 | 100 | 10 | 50 | 1443 | 0.9330 | 669.7 | 670.8 | 0.2% |
| | 100,000 | 144,270 | 100 | 10 | 50 | 2885 | 0.9659 | 340.6 | 341.2 | 0.2% |
| | 200,000 | 288,539 | 100 | 10 | 50 | 5771 | 0.9828 | 171.8 | 172.1 | 0.2% |
| | 300,000 | 432,809 | 100 | 10 | 50 | 8656 | 0.9885 | 114.9 | 115.2 | 0.3% |
| | 500,000 | 721,348 | 100 | 10 | 50 | 14427 | 0.9931 | 69.1 | 69.2 | 0.2% |
| | 1,000,000 | 1,442,695 | 100 | 10 | 50 | 28854 | 0.9965 | 34.6 | 34.6 | 0.0% |
| | 2,000,000 | 2,885,390 | 100 | 10 | 50 | 57708 | 0.9983 | 17.3 | 18.0 | 4.0% |
| | 3,000,000 | 4,328,085 | 100 | 10 | 50 | 86562 | 0.9988 | 11.5 | 11.9 | 3.1% |
| | 5,000,000 | 7,213,475 | 100 | 10 | 50 | 144270 | 0.9993 | 6.9 | 6.7 | -3.3% |
| | 10,000,000 | 14,426,950 | 100 | 10 | 50 | 288539 | 0.9997 | 3.5 | 3.5 | 1.0% |

Conditions:
N documents = 10000
N copies of each document = 1
Document size = 50MB
Simulation length = 100,000 hours = 10 metric years
Sector size = 1MB
Simulation repeated n=101 times with different seeds

# Appendix C

Random sampling for auditing **without replacement** yields better results than sampling **with replacement**. We did a few small simulations to verify this intuition, as shown below. In this graph, the differences are large enough to be obvious only at high error rates (i.e., low sector lifetimes).

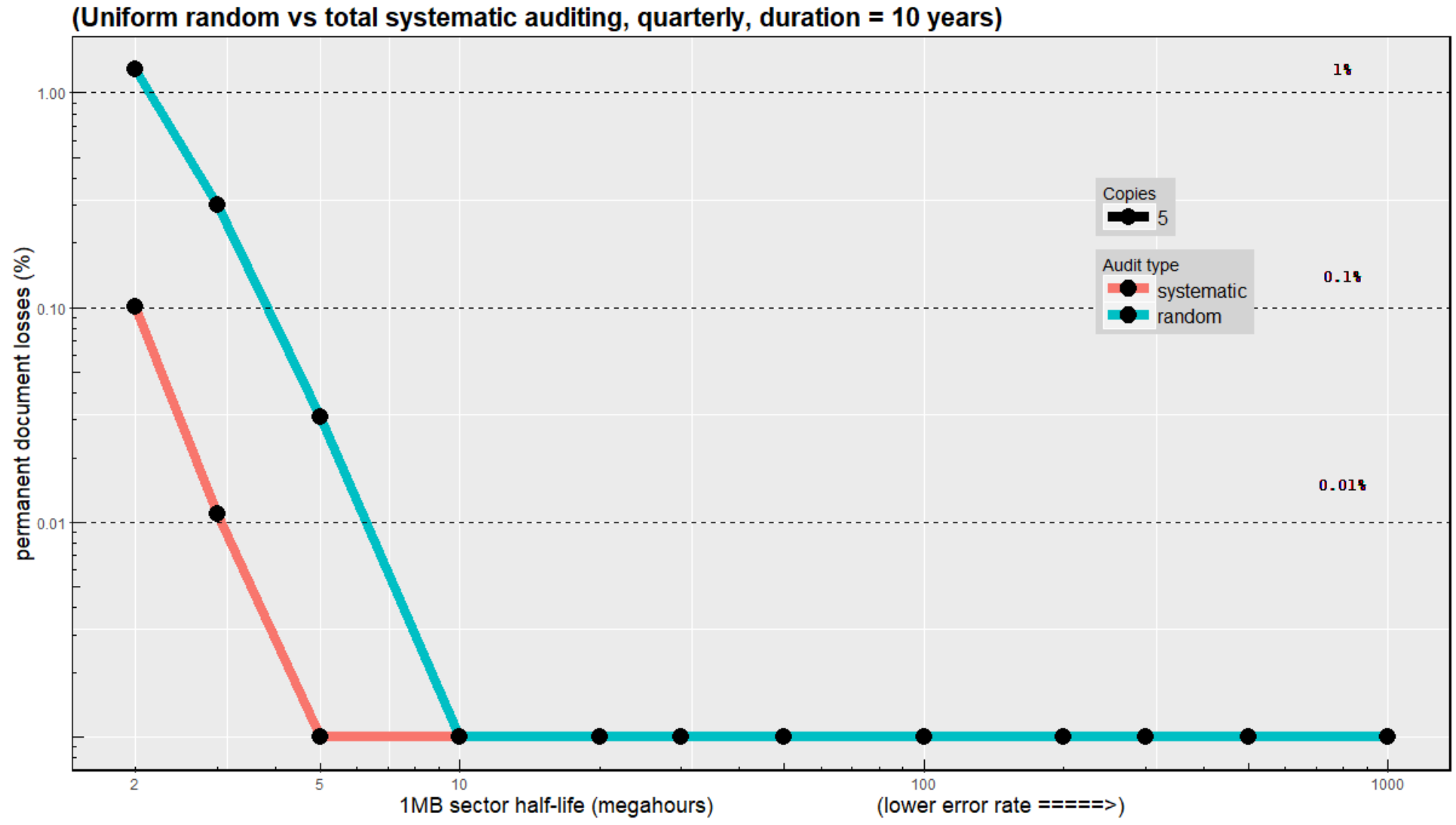